\documentclass[aps,prl,reprint,twocolumn,superscriptaddress,groupedaddress]{revtex4}

\usepackage{graphicx}% Include figure files
\usepackage{dcolumn}% Align table columns on decimal point
\usepackage{bm}% bold math
\usepackage{inputenc}
\usepackage{natbib}
\usepackage{graphicx}  % needed for figures
\usepackage{epsfig}
\usepackage{float}
\usepackage{dcolumn}   % needed for some tables
\usepackage{array}
\usepackage{amssymb}   % for math
\usepackage{amsmath}
\usepackage{amstext}
\usepackage{mathtools}
\usepackage{threeparttable}
\usepackage{siunitx} % for nicer Unit signs 
\usepackage{lipsum}  % if ever need blank text 
\usepackage{mathrsfs}
\usepackage{physics}
\usepackage{xcolor}
\usepackage[draft]{todonotes}

\usepackage [colorlinks=true,allcolors=blue]{hyperref}% add hypertext capabilities

\DeclareRobustCommand{\element}[1]{\@element#1\@nil}
\def\@element#1#2\@nil{%
  #1%
  \if\relax#2\relax\else\MakeLowercase{#2}\fi}
\makeatother

\begin{document}
\title{From Charge to Orbital Ordered Metal-Insulator Transition in Alkaline-Earth Ferrites}
\author{Yajun Zhang}
\affiliation{Department of Engineering Mechanics, School of Aeronautics and Astronautics, Zhejiang University, 38 Zheda Road, Hangzhou 310007, China}
\affiliation{Theoretical Materials Physics, Q-MAT, CESAM, Universit\'e de Li\`ege, Belgium}
\author{Michael Marcus Schmitt}
\affiliation{Theoretical Materials Physics, Q-MAT, CESAM, Universit\'e de Li\`ege, Belgium}
\author{Alain Mercy} 
\affiliation{Theoretical Materials Physics, Q-MAT, CESAM, Universit\'e de Li\`ege, Belgium}
\author{Jie Wang}
\email{jw@zju.edu.cn}
\affiliation{Department of Engineering Mechanics, School of Aeronautics and Astronautics, Zhejiang University, 38 Zheda Road, Hangzhou 310007, China}
\author{Philippe Ghosez}
\email{Philippe.Ghosez@ulg.ac.be}
\affiliation{Theoretical Materials Physics, Q-MAT, CESAM, Universit\'e de Li\`ege, Belgium}

\date{\today}% It is always \today, today,
             %  but any date may be explicitly specified
%%%%-----------------------------------------------------------------------%%%% 
%%%%                              ABSTRACT                                 %%%%
%%%%-----------------------------------------------------------------------%%%%
\begin{abstract}
%While CaFeO$_3$ exhibits upon cooling a metal-insulator transition at 290K, SrFeO$_3$ and BaFeO$_3$ keep a metallic behavior down to very low temperatures.
%Alkaline-earth ferrites show seemingly different behaviors. While CaFeO$_3$ exhibits upon cooling a metal-insulator transition linked to charge ordering similarly to rare-earth nickelates, SrFeO$_3$ and BaFeO$_3$ keep metallic behaviors down to very low temperatures. Moreover, opposite to rare-earth manganites, alkaline-earth ferrites do not seem prone to orbital ordering in spite of the identical $d^4$ formal occupancy of Fe$^{4+}$ and Mn$^{3+}$. 
While CaFeO$_3$ exhibits upon cooling a metal-insulator transition linked to charge ordering, SrFeO$_3$ and BaFeO$_3$ keep metallic behaviors down to very low temperatures. Moreover, alkaline-earth ferrites do not seem prone to orbital ordering in spite of the $d^4$ formal occupancy of Fe$^{4+}$.
Here, from first-principles simulations, we show that the metal-insulator transition of CaFeO$_3$ is structurally triggered by oxygen rotation motions as in rare-earth nickelates. This not only further clarifies why SrFeO$_3$ and BaFeO$_3$ remain metallic but allows us to predict that an insulating charge-ordered phase can be induced in SrFeO$_3$ from appropriate engineering of oxygen rotation motions. Going further, we unveil the possibility to switch from the usual charge-ordered to an orbital-ordered insulating ground state under moderate tensile strain in CaFeO$_3$ thin films. We rationalize the competition between charge and orbital orderings, highlighting alternative possible strategies to produce such a change of ground state, also relevant to manganite and nickelate compounds.
%%%%-----------------------------------------------------------------------%%%%
\end{abstract}

\pacs{Valid PACS appear here}% PACS, the Physics and Astronomy
                             % Classification Scheme.
%\keywords{Suggested keywords}%Use showkeys class option if keyword
                              %display desired
\maketitle

\begin{figure*}
	\includegraphics[width=\textwidth]{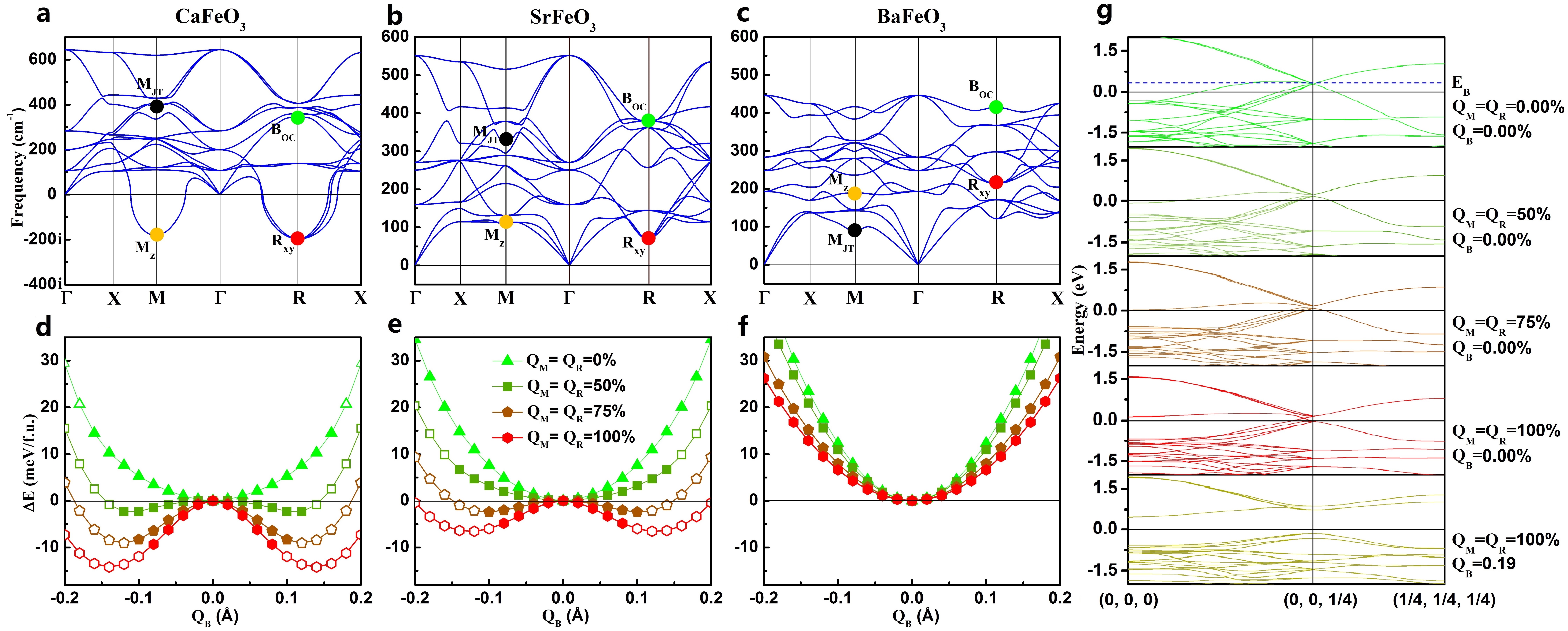}
    \caption{(a-c) Phonon dispersion curves of cubic CaFeO$_3$ (a), SrFeO$_3$ (b), and BaFeO3 (c) on which most relevant modes are pointed. (d-f) Evolution of the energy with respect to the breathing distortion amplitude ($Q_B$) at fixed rotation ($Q_M$) and tilts ($Q_R$) amplitudes in CaFeO$_3$ (d), SrFeO$_3$ (e), and BaFeO$_3$ (f). Opened (resp. filled) symbols denote insulating (resp. metallic) states. (g) Electronic band structure of CaFeO$_3$  along selected lines of the $Pbnm$ or $P2_1/n$ Brillouin zone (coordinates in pseudo-cubic notations) for different amplitude of disortions. All results were calculated with FM spin order and using a fixed cubic cell with the same volume as the ground state. Distortion amplitudes are normalized to those calculated by \textit{ISODISTORT}\cite{Campbell2006} in the CaFeO$_3$ AFM ground state.}
    \label{fig:Fig1_Phon_etc}
\end{figure*}
%%%%-----------------------------------------------------------------------%%%%

ABO$_3$ perovskite oxides, with a transition metal at the B-site, form a vast class of functional materials, fascinating by the diversity of their unusual properties \cite{khomskii2014transition,Imada1998,Zubko2011}. Amongst them, different families of compounds with a formal $e_g^1$ occupation of the \textit{d} orbitals at the B-site, like rare-earth manganites ($d^4 = t_{2g}^1 e_g^1$ in R$^{3+}$Mn$^{3+}$O$_3$, with R a rare-earth element), rare-earth nickelates ($d^7 = t_{2g}^6 e_g^1$ in R$^{3+}$Ni$^{3+}$O$_3$), or alkaline earth ferrites ($d^4 = t_{2g}^1 e_g^1$ in A$^{2+}$Fe$^{4+}$O$_3$, with A $=$ Ca, Sr or Ba) are similarly prone to show metal-insulator transitions (MIT). However, the mechanism behind such a transition can be intriguingly different from one family to the other. 

RNiO$_3$ (except R=La) and RMnO$_3$ compounds crystallize in the same metallic \textit{Pbnm} GdFeO$_3$-type phase at sufficiently high-temperature. This phase is compatible with their small tolerance factor and labeled ($a^-a^-c^+$) in Glazer’s notation\cite{Glazer1972}. It differs from the aristotype cubic perovskite structure, only expected at very high temperature and not experimentally observed, by the coexistence of two types of atomic distortions: (i) in-phase rotation of the oxygen octahedra along $z$ direction (M$_z$) and (ii) anti-phase tilts of the same oxygen octahedra with identical amplitude around $x$ and $y$ directions (R$_{xy}$). On the one hand, RNiO$_3$ compounds show on cooling a MIT ($T_{MIT} = 0-600K$) concomitant with a structural transition from \textit{Pbnm} to $P2_{1}/n$ \cite{Medarde1997}. This lowering of symmetry arises from the appearance of a breathing distortion of the oxygen octahedra (B$_{oc}$), recently assigned to a structurally triggered mechanism \cite{Mercy2017} and producing a kind of charge ordering (CO)\cite{Mazin2007,Park2012,Johnston2014,Varignon2017a}. On the other hand, RMnO$_3$ compounds also exhibit on cooling a MIT ($T_{MIT} \approx 750K$) but associated to orbital ordering (OO) and linked to the appearance of Jahn-Teller distortions (M$_{JT}$) compatible with the $Pbnm$ symmetry \cite{Kimura2003,Sanchez2003}. 

In comparison, AFeO$_3$ compounds do not behave so systemically and adopt seemingly different behaviors. While SrFeO$_3$ and BaFeO$_3$ keep the ideal cubic perovskite structure and show metallic behavior at all temperatures \cite{MacChesney1965,Hayashi2011}, CaFeO$_3$, which crystallizes above room temperature in a $Pbnm$ GdFeO$_3$-type phase, exhibits a behavior similar to nickelates. At 290K, a MIT takes place at the same time as its symmetry is lowered to $P2_1/n$ due to the appearance of a breathing distortion \cite{kawasaki1998phase,Woodward2000}. A variety of explanations have been previously proposed to elucidate the MIT in CaFeO$_3$ , including orbital hybridization \cite{Akao2003}, electron-lattice interactions \cite{Matsuno2002,Ghosh2005}, and ferromagnetic coupling \cite{Cammarata2012}. However, no net picture has emerged yet to rationalize its behavior and that of other ferrites. 

Here, we show from first-principles calculations that the CO-type MIT in bulk CaFeO$_3$ arises from the same microscopic mechanism as in the nickelates and must be assigned to a progressive triggering of B$_{oc}$ atomic distortions by M$_z$ and R$_{xy}$ atomic motions. We demonstrate that this triggered mechanism is universal amongst the ferrite family and that an insulating phase can be induced in metallic SrFeO$_3$ from appropriate tuning of oxygen rotations and tilts. Going further, we reveal that CO and OO compete in AFeO$_3$ compounds and we unveil the possibility to switch from CO-type to OO-type MIT in CaFeO$_3$ thin films under appropriate strain conditions. This offers a convincing explanation for the enormous resistivity at room-temperature recently found in CaFeO$_3$ films grown on SrTiO$_3$ \cite{Rogge2018}.

\emph{Methods} - Our first-principles calculations relied on density functional theory (DFT) as implemented in VASP \cite{Kresse1993,Bloechl1994}. We worked with the PBEsol \cite{Perdew2008} exchange-correlation functional including U and J corrections as proposed by Liechtenstein \cite{Anisimov1997}. We used $(U|J) = (7.2|2.0) \si{\eV}$, a plane-wave energy cutoff of $600 eV$ and Monkhorst-Pack\cite{PhysRevB.13.5188} k-point samplings equivalent to $12 \times 12 \times 12$ for a 5-atoms cubic perovskite cell. The lattice parameters and internal atomic coordinates were relaxed until atomic forces are less than $10^{-5}\si{\eV}/\si{\angstrom}$. The phonon dispersion curves were calculated with  $2 \times 2 \times 2$  supercells using finite displacement method. A special care was devoted to the determination of appropriate U and J parameters, which is discussed in detail in the Supplementary Material (SM) \cite{foot}. We found that  $(U|J) = (7.2|2.0) \, \si{\eV}$ provides good simultaneous description of the structural (lattice constant and distortion amplitudes), electronic (insulating ground- state) and magnetic (AFM spiral-type ground state very close in energy to the FM configuration) properties of CaFeO$_3$.
\emph{Bulk CaFeO$_3$ --} In order to clarify the mechanism behind the $P2_1/n$ insulating ground state of CaFeO$_3$, we first focus on the phonon dispersion curves of its parent cubic phase (Fig. \ref{fig:Fig1_Phon_etc}a). Calculations are reported in a ferromagnetic configuration, which is representative to unravel the essential physics. On the one hand, Fig. \ref{fig:Fig1_Phon_etc}a) shows expected unstable phonon modes at M point ($M_2^+$, $\omega_M=181i \si{\cm}^{-1}$) and R point ($R_5^-$, $\omega_R=197i \si{\cm}^{-1}$) of the Brillouin zone, related respectively to the M$_z$ and R$_{xy}$ distortions yielding the \textit{Pbnm} phase. On the other hand, it attests that the $R_2^-$ mode related to the B$_{oc}$ distortion is significantly stable ($\omega_B^2 = 343\si{cm}^{-1}$), so questioning the origin of its appearance in the $P2_1/n$ phase. 

The answer is provided in Fig. \ref{fig:Fig1_Phon_etc}b), reporting the evolution of the energy with the amplitude of B$_{oc}$ ($Q_B$) at fixed amplitudes of M$_z$ ($Q_M$) and R$_{xy}$ ($Q_R$). It demonstrates that, although initially stable (single well -- SW -- with a positive curvature at the origin $\alpha_B \propto\omega_B^2 > 0$) in the cubic phase, B$_{oc}$ will be progressively destabilized (double well -- DW -- with a renormalized negative curvature at the origin $\tilde{\alpha}_B< 0$) as M$_z$ and R$_{xy}$ develop in the \textit{Pbnm} phase. The curvature $\tilde{\alpha}_B$ changes linearly with $Q_M^2$ and $Q_R^2$ ($\tilde{\alpha}_B = \alpha_B + \lambda_{BM} Q_M^2 + \lambda_{BR} Q_R^2$) so that its evolution must be assigned to a \textit{cooperative} biquadratic coupling ($\lambda_{BM}, \lambda_{BR}<0$) of $B_{oc}$ with M$_z$ and R$_{xy}$ as highlighted by the following terms in the Landau-type energy expansion around the cubic phase: 
\begin{equation}
	E \propto \alpha_B Q_B^2 + \lambda_{BM} Q_M^2 Q_B^2 + \lambda_{BR} Q_R^2 Q_B^2
\end{equation}
For large enough amplitudes of M$_z$ and R$_{xy}$, B$_{oc}$ becomes unstable and will spontaneously appear in the structure. In Fig. \ref{fig:Fig1_Phon_etc}b) we further notice that the amplitude of B$_{oc}$ required for making the system insulating decreases for increasing M$_z$ and R$_{xy}$, yielding therefore an insulating $P2_1/n$ ground state. 

This behavior is point by point similar to that reported recently in rare-earth nickelates by Mercy \textit{et al.}\cite{Mercy2017} who subsequently assigned the MIT to a structurally triggered phase transition, in the sense originally defined by Holakovsk\'{y} \cite{Holakovsky1973}. In Ref. \cite{Mercy2017}, the unusual cooperative coupling of B$_{oc}$ with M$_z$ and R$_{xy}$ at the origin of this triggered mechanism was moreover traced back in the electronic properties of nickelates and further related to a type of structurally triggered Peierls instability. 

Fig \ref{fig:Fig1_Phon_etc}g) shows that this explanation still hands for ferrites. In the cubic phase of CaFeO$_3$ (Fe$^{4+}$ with formal occupation $d^4 = t_{2g}^1 e_g^1$), the Fermi energy, E$_F$, crosses anti-bonding Fe $3d$ -- O $2p$ states with a dominant $e_g$ character. Activation of B$_{oc}$ can open a gap in these partly occupied $e_g$ bands at $q_B=(1/4,1/4,1/4)$, but around an energy E$_B$ initially above E$_F$. The role of R$_{xy}$ and $M_{z}$ is to tune Fe $3d$ -- O $2p$ hybridizations in such a way that E$_B$ is progressively lowered towards E$_F$.  As they develop into the structure, activating B$_{oc}$ affects more and more substantially energy states around E$_F$ and yields an increasing gain of electronic energy explaining the progressive softening of $\omega_B$. The $e_g$ bandwidth in CaFeO$_3$ being smaller than in the nickelates, E$_B$ is initially closer to E$_F$ consistently with a softer $\omega_B$ and the smaller amplitude of R$_{xy}$ and M$_{z}$ required to destabilize B$_{oc}$. 

%%%%-----------------------------------------------------------------------%%%%
%%%%                       SECTION BULK SFO AND BFO                        %%%%
%%%%-----------------------------------------------------------------------%%%%
%\section{\label{sec:B_SFO_BFO} Bulk \element{Sr}\element{Fe}O$\mathbf{_3}$ and \element{Ba}\element{Fe}O$\mathbf{_3}$}
\emph{Bulk SrFeO$_3$ and BaFeO$_3$ --} The triggered mechanism highlighted above further straightforwardly explains the absence of MIT in other alkaline-earth ferrites. Because of their larger tolerance factors and as confirmed from the absence of unstable mode in their phonon dispersion curves (Fig. \ref{fig:Fig1_Phon_etc}b) and c)) SrFeO$_3$  and BaFeO$_3$  preserve their cubic structure down to zero Kelvin\cite{MacChesney1965,Hayashi2011} and so do not spontaneously develop the oxygen rotation and tilts mandatory to trigger the MIT. The cooperative coupling of B$_{oc}$ with M$_z$ and R$_{xy}$ remains however a generic features of all ferrite compounds. 

As illustrated in Fig.  \ref{fig:Fig1_Phon_etc}e) and \ref{fig:Fig1_Phon_etc}f), B$_{oc}$ is progressively destabilized when increasing artificially the amplitudes of M$_z$ and R$_{xy}$ distortions in SrFeO$_3$  and BaFeO$_3$. Since, in the cubic phase of these compounds, $\omega_B$ is originally at frequencies slightly larger than in CaFeO$_3$  ($\omega_B =  362 \si{cm}^{-1}$ in SrFeO$_3$ and $\omega_B =  415 \si{cm}^{-1}$ in BaFeO$_3$), larger distortions are required to induce the MIT. In SrFeO$_3$, amplitudes of M$_z$ and R$_{xy}$ corresponding to 75\% of their ground-state values in CaFeO$_3$ are nevertheless enough to force an insulating ground state. In BaFeO$_3$, the cooperative coupling is less efficient and much larger amplitudes would be required. 

This highlights the possibility of inducing a MIT in SrFeO$_3$ thin films or heterostructures under appropriate engineering of R$_{xy}$ and M$_{z}$. Moreover, it provides a vivid explanation to the decrease of T$_{MIT}$ experimentally observed in Ca$_{1-x}$Sr$_x$FeO$_3$ solid solutions as $x$ increases \cite{Takeda2000}. For increasing Sr concentrations, the average tolerance factor increases and the mean  amplitudes of M$_z$ and R$_{xy}$ decrease. This analysis is supported by DFT calculation at 50/50 Ca/Sr composition using an ordered supercell (see SM \cite{foot}). 

%%%%-----------------------------------------------------------------------%%%%
%%%%                      CHARGE VS. ORBITAL ORDERING                      %%%%
%%%%-----------------------------------------------------------------------%%%%
%\section{\label{sec:CO_VS_OO} Charge Order vs. Orbital Ordering}
\emph{Charge versus orbital ordering --} %The mechanism of the MIT being clarified, 
It remains intriguing why CaFeO$_3$  ($t_{2g}^3e_g^1$) prefers to exhibit a breathing distortion (B$_{oc}$) and CO as RNiO$_3$  compounds ($t_{2g}^6e_g^1$) rather than a Jahn-Teller distortion (M$_{JT}$) and OO as RMnO$_3$ compounds ($t_{2g}^3e_g^1$). In Ref. \cite{Whangbo2002} Whangbo \textit{et al.} argue that B$_{oc}$ is favored in CaFeO$_3$ by the relatively strong covalent character of the Fe-O bond while the M$_{JT}$ distortion is preferred in LaMnO$_3$ by the weak covalent character of the Mn-O bond. So, we anticipate that weakening the covalence by increasing the Fe-O distance might favor M$_{JT}$ and OO in CaFeO$_3$. To realize practically this idea, we investigated the role of tensile epitaxial strain on the ground state of CaFeO$_3$  thin films.

%%%%-----------------------------------------------------------------------%%%%
%%%%                             CFO THIN FILMS                            %%%%
%%%%-----------------------------------------------------------------------%%%%
%\section{\label{sec:CO_Thin_Film} \element{Ca}\element{Fe}O$_\mathbf{3}$ Thin films}
%%%%-----------------------------------------------------------------------%%%% 
%%%%                      FIGURE 2 STRAIN PHASE DIAGRAM                    %%%%
%%%%-----------------------------------------------------------------------%%%%
\begin{figure}
	\includegraphics[width=\columnwidth]{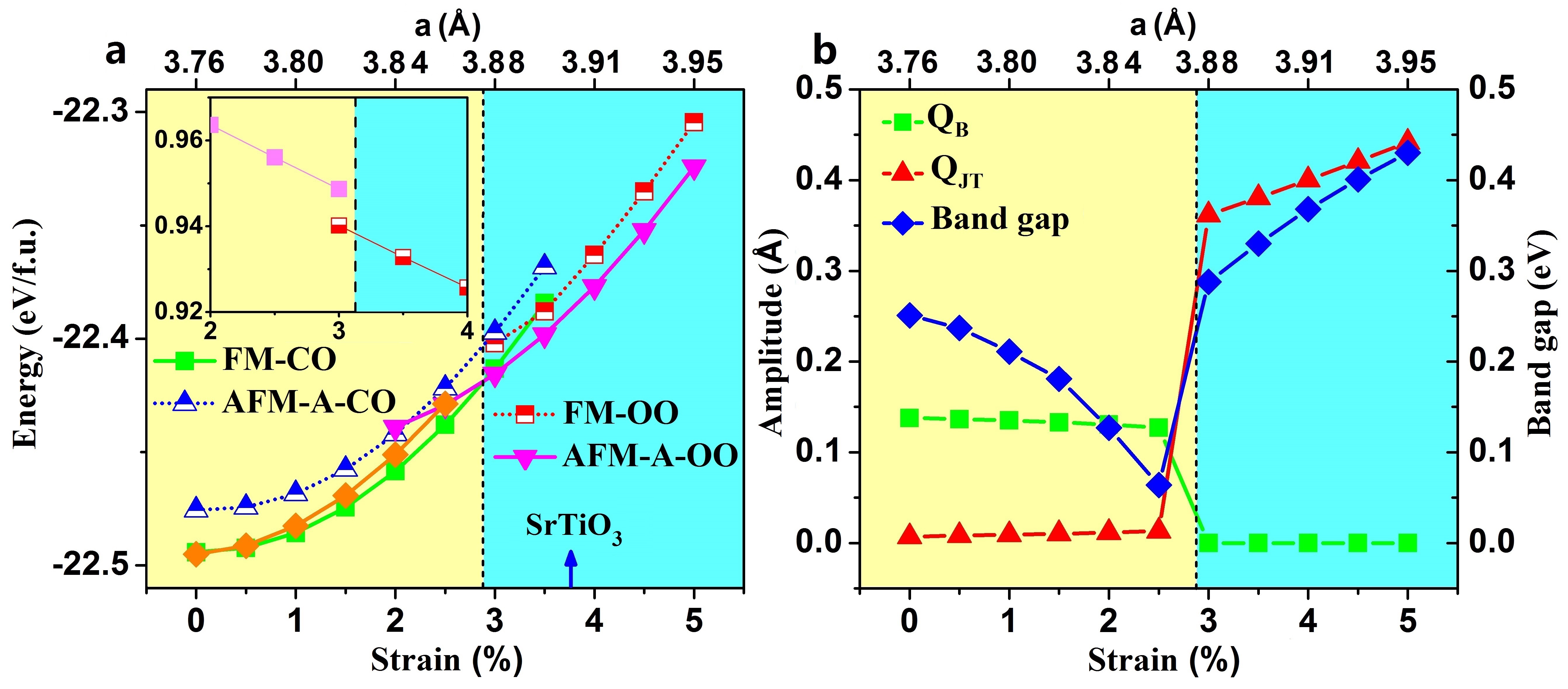}
    \caption{(a) Total energy as a function of tensile strain (or in-plane lattice constant) for CaFeO$_3$ epitaxial films with c-axis out-of-plane, in CO and OO states with either FM or A-type AFM spin ordering; FM CO state with the long c-axis in-plane is plot in orange for comparison. A change of ground state from FM-CO phase (yellow area) to A-type AFM-OO phase (blue area) is observed. Inset: c/a ratio of the ground state structure as a function of strain. (b) Evolution of $Q_{B}$ (green), $Q_{JT}$ (red) and band gap (blue) as a function of strain (or in-plane lattice constant).}
    \label{fig:Fig3_Phase Diagram.}
\end{figure}
%%%%-----------------------------------------------------------------------%%%%
%%%%-----------------------------------------------------------------------%%%% 
%%%%                        FIGURE 3 JT-COUPLINGS ETC                       %%%%
%%%%-----------------------------------------------------------------------%%%%
\begin{figure*}
	\includegraphics[width=0.8\textwidth]{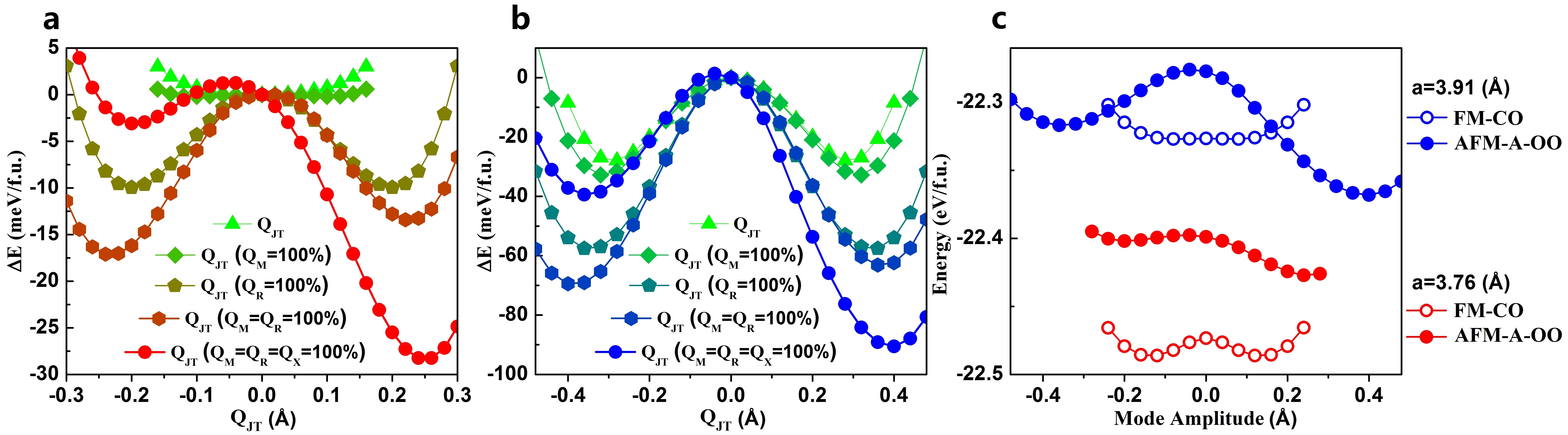}
    \caption{ Evolution of the energy with Jahn-Teller distortion amplitude $Q_{JT}$ in AFM-A magnetic order and at fixed amplitudes of other distortion (see legend) for CaFeO$_3$ epitaxial films under strain of (a) 0\% ($a=3.76$ \si{\angstrom}) and (b) 4\% ($a=3.91$ \si{\angstrom}). %(c) Evolution of the total energy as a function of $Q_B$ in FM-CO state (open symbol) and $Q_{JT}$ in A-type AFM-OO state (filled symbol) for CaFeO$_3$ thin film under 0\% ($a=3.76$\si{\angstrom}, in red open symbol $Q_M$, $Q_R$ and $Q_X$ of FM-CO phase is fixed, the corresponding mode amplitude of A-type AFM-OO is fixed for the filled symbol) and 4\% ($a=3.91$\si{\angstrom}, blue curves with fixed $Q_M$, $Q_R$ and $Q_X$ of of A-type AFM-OO phase, since the initial CO phase converges to OO phase after optimization under such strain) strains.
(c) Evolution of the total energy as a function of $Q_B$ in FM-CO state (open symbol) and $Q_{JT}$ in A-type AFM-OO state (filled symbol) for CaFeO$_3$ thin film under 0\% ($a=3.76$\si{\angstrom}, red) and 4\% ($a=3.91$\si{\angstrom}, blue) strains. $Q_M$, $Q_R$ and $Q_X$  are fixed to their amplitudes in the relevant phase  (except for the FM-CO state at 4\% which cannot be stabilized and for which we kept positions in the  A-type AFM-OO phase).
}
    \label{fig:Fig4:JT_couplings}
\end{figure*}
%%%%-----------------------------------------------------------------------%%%%

\emph{CaFeO$_3$ thin films --} The phase diagram of CaFeO$_3$  films epitaxially grown on a cubic perovskite (001)-substrate is reported Fig. \ref{fig:Fig3_Phase Diagram.}a). 
The evolution of the energy with the lattice constant of the substrate is shown for FM and A-type AFM orders with either charge or orbital ordering. Although S- and T-type spiral magnetic orders (not shown here) possess a slightly lower energy at the bulk level, the FM order becomes quickly the GS under small tensile strain; C-type and G-type AFM order are much higher in energy and not shown. Both possible orientations of the orthorhombic ($a^-a^-c^+$) oxygen rotation pattern, with the long c-axis either in-plane or out-of-plane were also considered: while c-axis in-plane is favored at zero strain, c-axis out-of-plane becomes more stable under tensile strains. 

Fig. \ref{fig:Fig3_Phase Diagram.}a) demonstrates the possibility of switching from a CO to an OO ground state in CaFeO$_3$  using strain engineering:  under increasing tensile strain, the ground state of the film changes from an insulating FM-CO $P2_1/n$ configuration at small strain to an insulating A-type AFM-OO \textit{Pbnm} configuration above 3\% tensile strain (a=3.88\si{\angstrom}). Fig. \ref{fig:Fig3_Phase Diagram.}b) highlights the strain evolution of B$_{oc}$ and M$_{JT}$ distortions together with the change of band gap. Under increasing tensile strain, B$_{oc}$ slightly decreases and is abruptly suppressed at the phase transition; at the same time, the band gap -- already reduced in this FM phase -- decreases, although much faster than B$_{oc}$ and the transition appears precisely when the bandgap converges to zero. Conversely, M$_{JT}$ is nearly zero below 3\% tensile strain while it suddenly appears at the transition and then continuously increases.  Amazingly, the amplitude of M$_{JT}$ (0.37 \si{\angstrom}) in a CaFeO$_3$  film grown on a SrTiO$_3$  substrate (a = 3.905 \si{\angstrom}) is comparable to that of bulk LaMnO$_3$  (0.36 \si{\angstrom}). Such similar amplitude suggests that the $T_{MIT}$ associated to the OO state in strained CaFeO$_3$  films might be much larger than the $T_{MIT}$ associated to the CO state in bulk and comparable to the one of LaMnO$_3$  ($T_{MIT}$ = 750K). 

Our findings provide a convincing explanation for the insulating character of CaFeO$_3$ films on SrTiO$_3$ at room temperature and for the absence of CO MIT in the 100-300K temperature range as recently pointed out in Ref.\cite{Rogge2018}. They suggest to probe the presence of OO MIT at higher temperature. A key feature, highlighted in the insert in Fig. \ref{fig:Fig3_Phase Diagram.}a), is the jump of c/a ratio at the transition boundary, which provides another concrete hint for experimentalists to probe the CO-OO transition.
%%%%-----------------------------------------------------------------------%%%%
%%%%                      COMPETITION BETWEEN CO AND OO                    %%%%
%%%%-----------------------------------------------------------------------%%%%
%\section{\label{sec:Comp_CO_OO} Competition between Charge and Orbital Ordering}

%%%%-----------------------------------------------------------------------%%%% 
%%%%                      TABLE FITTING PARAMETERS                         %%%%
%%%%-----------------------------------------------------------------------%%%%
\begin{table}
\renewcommand{\arraystretch}{1.35}
 \caption{Top: Amplitudes of dominant distortions \cite{Campbell2006} in the relaxed CO (FM) and OO (AFM-A) phases of CaFeO$_3$ epitaxial films under 0\% ($a=3.76$\si{\angstrom}) and 4\% ($a=3.91$\si{\angstrom}) tensile strain. Bottom : Energy contributions associated to the different terms in Eq.\eqref{eq:E_JT}, obtained from the amplitudes of distortion reported above.}
    \label{tab:Tab2_Fitting_Paras}
	\begin{ruledtabular}
			\begin{tabular*}{\columnwidth}{ccccccc}
     	Amplitudes (\si{\angstrom}) & $Q_{JT}$ & $Q_B$ & $Q_M$ & $Q_R$ & $Q_X$ \\
           \hline 
       a = 3.76\si{\angstrom}-CO & 0.007 & 0.137  & 0.788 &  1.111 & 0.440  \\
       a = 3.76\si{\angstrom}-OO & 0.256 & 0.000  & 0.717 &  1.194 & 0.457  \\
       a = 3.91\si{\angstrom}-OO & 0.400 & 0.000 & 0.676 &  1.309 & 0.552  \\
     %   \hline
      %   Parameters & $\alpha_{JT}$ & $\lambda_{MJT}$ & $\lambda_{RJT}$ & \multicolumn{2}{l}{$\gamma_{RXJT}$} \\         \hline 
       %a = 3.76\si{\angstrom}-OO & 37.4 & -196  & -338.3 &  \multicolumn{2}{l}{$5.2$}  \\
      % a = 3.91\si{\angstrom}-OO & -672.4 & -41.2 & -256.5  & \multicolumn{2}{l}{$107.2$}  \\
           \hline
        Energies (\si{\meV}/fu) & $\Delta E_{JT}^{(1)}$ & $\Delta E_{JT}^{(2)}$  & $\Delta E_{JT}^{(3)}$  & $\Delta E_{JT}$ & $\Delta E_{AFM-A}$  \\         \hline 
       a = 3.76\si{\angstrom}-OO & 4.4 & -43.6  & -15.8&  -55.0 & 74.5  \\
       a = 3.91\si{\angstrom}-OO & -104.3 & -80.9 & -31.0  & -216.2 & 49.3  \\
            \end{tabular*}
    \end{ruledtabular}
\end{table}
%%%%-----------------------------------------------------------------------%%%%

\emph{Competition between charge and orbital orders} - To rationalize the emergence of an OO ground state in CaFeO$_3$ films, we quantify the lowest-order couplings of M$_{JT}$ with other distortions in a Landau-type free energy expansion and investigate their sensitivity to magnetic order and epitaxial strain :
\begin{align}
	\centering
	E \propto \alpha_{JT} Q_{JT}^2 + & \lambda_{MJ} Q_M^2 Q_{JT}^2 + \lambda_{RJ} Q_R^2 Q_{JT}^2  \nonumber \\ 
       & + \gamma  Q_R Q_X Q_{JT} 
    \label{eq:E_JT}
\end{align}
The first term quantifies the \textit{proper} harmonic energy contribution $\Delta E_{JT}^{(1)}$ associated to the appearance of M$_{JT}$. The second and third terms in Eq. \eqref{eq:E_JT} account for a change of energy $\Delta E_{JT}^{(2)}$ in presence of M$_z$ and R$_{xy}$, linked to their lowest bi-quadratic coupling with M$_{JT}$. Finally, sizable anti-polar motions of the Ca cations and apical oxygens (X$_{AP}$ mode of amplitude $Q_X$, see Table \ref{tab:Tab2_Fitting_Paras}), which are driven by M$_z$ and R$_{xy}$\cite{Varignon2015}, couple in a trilinear term with R$_{xy}$ and M$_{JT}$ (last term in Eq.\eqref{eq:E_JT}). This coupling produces an energy lowering $\Delta E_{JT}^{(3)} <0$, through a so-called \textit{hybrid improper} mechanism yeilding an asymmetry in the M$_{JT}$ energy well \footnote{The coexistence of M$_z$ and R$_{xy}$ already produces an asymmetry in the energy well through the term $E\propto \delta Q_R^2 Q_M Q_{JT}$. However, this asymmetry is in the negligible range of 1 \si{\meV} (see Fig. \ref{fig:Fig4:JT_couplings}a-b), and is not further discussed here.}. Compendiously, appearance of a M$_{JT}$ distortion requires $\Delta E_{JT} =  \Delta E_{JT}^{(1)} + \Delta E_{JT}^{(2)} + \Delta E_{JT}^{(3)} < 0$.

In bulk CaFeO$_3$, $\alpha_{JT}$ is large ($\omega_{JT} (FM)= 390$ cm$^{-1}$) in the FM cubic phase, which prohibits $\Delta E_{JT}$ to become negative for sizable amplitudes of M$_{JT}$. Switching to the A-type AFM spin order tremendously lowers $\alpha_{JT}$ ($\omega_{JT} (AFM-A)= 144$ cm$^{-1}$) but simultaneously increases the total energy by $\Delta E_{AFM-A}$. The stabilization of an OO phase with M$_{JT}$ against the CO phase with B$_{oc}$ so depends eventually on the counterbalance between $\Delta E_{AFM-A}$ and $\Delta E_{JT}$.

This is quantified for epitaxial thin films in Fig. \ref{fig:Fig4:JT_couplings} and Table \ref{tab:Tab2_Fitting_Paras}. Under negligible tensile strain (a= 3.76 \AA, Fig. \ref{fig:Fig4:JT_couplings}a), with A-type AFM order, $\omega_{JT}$ is even softer than in bulk CaFeO$_3$, yielding $\Delta E_{JT}^{(1)} \approx 0$. Then, similarly to what was discussed for $B_{oc}$ in bulk compounds, $R_{xy}$ and $M_z$ trigger $M_{JT}$ ($\lambda_{MJ},\lambda_{RJ}<0$), yielding $\Delta E_{JT}^{(2)} <0$. Finally, the hybrid improper coupling with $X_{AP}$ and R$_{xy}$ provides a further $\Delta E_{JT}^{(3)} <0$. However, although globally negative, $\Delta E_{JT}$ cannot overcome $\Delta E_{AFM-A}$(Fig. \ref{fig:Fig4:JT_couplings}c). Under large tensile strain (a= 3.91 \AA, Fig. \ref{fig:Fig4:JT_couplings}b) $\alpha_{JT}$ is significantly reduced by coupling with the epitaxial tetragonal strain $e_{tz}$ ($\alpha_{JT} \propto  \gamma_{tJ}e_{tz} + \lambda_{tJ} e_{tz}^2$ \cite{Carpenter2009}), yielding a huge negative $\Delta E_{JT}^{(1)}$. Then, although $\lambda_{MJ}$ and $\lambda_{RJ}$ are reduced and $\gamma$ remains unaffected (see SM \cite{foot}), $\Delta E_{JT}^{(2)}$ and $\Delta E_{JT}^{(3)}$ are increased roughly by a factor of 2, mainly due to the increase of $Q_{JT}$. Globally, $|\Delta E_{JT}|$ in A-type AFM order is now much larger than $\Delta E_{AFM-A}$, which moreover has been slightly reduced, and the OO phase with M$_{JT}$ is stabilized against the CO phase with B$_{oc}$ (Fig. 3c). We notice that $Q_M$ and $Q_R$ are not strongly affected by strain so that the stabilization of the OO phase must be primarily assigned to the strain remormalization of $\alpha_{JT}$. $E_{JT}^{(2)}$ and $\Delta E_{JT}^{(3)}$ play however an important complementary role and tuning  $Q_M$ and $Q_R$ would offer an alternative strategy to stabilize the OO phase. 
%Further, we highlight that the asymmetry of energy well induced by $\Delta E_{JT}^{(3)}$ originates from the competition between $ Q_X^A Q_R Q_{JT}$ and $Q_X^O Q_R Q_{JT}$ couplings. $Q_X^O$ denotes the oxygen movement and $Q_X^A$  is the cation movement appearing in the same irreducible representation X5- and this two motions make opposite contribution to $\Delta E_{JT}^{(3)}$. From CO phase to OO phase, $Q_X$ is significantly increased leading to the substantial change of asymmetry (see Supplementary Materials)

%%%%-----------------------------------------------------------------------%%%%
%%%%                             CONCLUSIONS                               %%%%
%%%%-----------------------------------------------------------------------%%%%
\emph{Conclusions} - We have rationalized the appearance of a CO-type MIT in alkaline-earth ferrites, showing that, in CaFeO$_3$, such a MIT arises from the triggering of $B_{oc}$ by M$_z$ and R$_{xy}$ and that this mechanism can induce a CO insulating ground state in SrFeO$_3$ under appropriate tuning of M$_z$ and R$_{xy}$. Going further, we found that OO is also incipient to CaFeO$_3$ and that an OO-type MIT can be engineered in thin films under moderate tensile strain. We have shown that the appearance of the OO-type insulating ground state arises from a delicate balance between different energy terms, suggesting different strategies to stabilize it. Interestingly, the emergence of the OO phase in ferrites is the result of a purely structural instability and we did not find any gradient discontinuity in the energy (corner point), fingerprint of the electronic instability usually associated to OO phases\cite{Yin2006}. Such a structural stabilization of the OO phase might offer a reasonable explanation to the  emergence of an OO phase in other materials like RNiO$_3$ compounds \cite{He2015}\footnote{Xu He et.al, in press.}.

\emph{Acknowledgments} - Work supported by the FRS-FNRS PDR project HiT4FiT and ARC AIMED. M.S. and Y. Z. acknowledge financial support from FRIA (grants 1.E.070.17. and 1.E.122.18.). Computational support from C\'eci funded by F.R.S-FNRS (Grant No. 2.5020.1) and Tier-1 supercomputer of the F\'ed\'eration Wallonie-Bruxelles funded by the Walloon Region (Grant No. 1117545). M.S. and Y.Z. contributed equally to this work. 


\begin{thebibliography}{34}%
\makeatletter
\providecommand \@ifxundefined [1]{%
 \@ifx{#1\undefined}
}%
\providecommand \@ifnum [1]{%
 \ifnum #1\expandafter \@firstoftwo
 \else \expandafter \@secondoftwo
 \fi
}%
\providecommand \@ifx [1]{%
 \ifx #1\expandafter \@firstoftwo
 \else \expandafter \@secondoftwo
 \fi
}%
\providecommand \natexlab [1]{#1}%
\providecommand \enquote  [1]{``#1''}%
\providecommand \bibnamefont  [1]{#1}%
\providecommand \bibfnamefont [1]{#1}%
\providecommand \citenamefont [1]{#1}%
\providecommand \href@noop [0]{\@secondoftwo}%
\providecommand \href [0]{\begingroup \@sanitize@url \@href}%
\providecommand \@href[1]{\@@startlink{#1}\@@href}%
\providecommand \@@href[1]{\endgroup#1\@@endlink}%
\providecommand \@sanitize@url [0]{\catcode `\\12\catcode `\$12\catcode
  `\&12\catcode `\#12\catcode `\^12\catcode `\_12\catcode `\%12\relax}%
\providecommand \@@startlink[1]{}%
\providecommand \@@endlink[0]{}%
\providecommand \url  [0]{\begingroup\@sanitize@url \@url }%
\providecommand \@url [1]{\endgroup\@href {#1}{\urlprefix }}%
\providecommand \urlprefix  [0]{URL }%
\providecommand \Eprint [0]{\href }%
\providecommand \doibase [0]{http://dx.doi.org/}%
\providecommand \selectlanguage [0]{\@gobble}%
\providecommand \bibinfo  [0]{\@secondoftwo}%
\providecommand \bibfield  [0]{\@secondoftwo}%
\providecommand \translation [1]{[#1]}%
\providecommand \BibitemOpen [0]{}%
\providecommand \bibitemStop [0]{}%
\providecommand \bibitemNoStop [0]{.\EOS\space}%
\providecommand \EOS [0]{\spacefactor3000\relax}%
\providecommand \BibitemShut  [1]{\csname bibitem#1\endcsname}%
\let\auto@bib@innerbib\@empty
%</preamble>
\bibitem [{\citenamefont {Khomskii}(2014)}]{khomskii2014transition}%
  \BibitemOpen
  \bibfield  {author} {\bibinfo {author} {\bibfnamefont {D.}~\bibnamefont
  {Khomskii}},\ }\href@noop {} {\emph {\bibinfo {title} {Transition metal
  compounds}}}\ (\bibinfo  {publisher} {Cambridge University Press},\ \bibinfo
  {year} {2014})\BibitemShut {NoStop}%
\bibitem [{\citenamefont {Imada}\ \emph {et~al.}(1998)\citenamefont {Imada},
  \citenamefont {Fujimori},\ and\ \citenamefont {Tokura}}]{Imada1998}%
  \BibitemOpen
  \bibfield  {author} {\bibinfo {author} {\bibfnamefont {M.}~\bibnamefont
  {Imada}}, \bibinfo {author} {\bibfnamefont {A.}~\bibnamefont {Fujimori}}, \
  and\ \bibinfo {author} {\bibfnamefont {Y.}~\bibnamefont {Tokura}},\ }\href
  {\doibase 10.1103/RevModPhys.70.1039} {\bibfield  {journal} {\bibinfo
  {journal} {Rev. Mod. Phys.}\ }\textbf {\bibinfo {volume} {70}},\ \bibinfo
  {pages} {1039} (\bibinfo {year} {1998})}\BibitemShut {NoStop}%
\bibitem [{\citenamefont {Zubko}\ \emph {et~al.}(2011)\citenamefont {Zubko},
  \citenamefont {Gariglio}, \citenamefont {Gabay}, \citenamefont {Ghosez},\
  and\ \citenamefont {Triscone}}]{Zubko2011}%
  \BibitemOpen
  \bibfield  {author} {\bibinfo {author} {\bibfnamefont {P.}~\bibnamefont
  {Zubko}}, \bibinfo {author} {\bibfnamefont {S.}~\bibnamefont {Gariglio}},
  \bibinfo {author} {\bibfnamefont {M.}~\bibnamefont {Gabay}}, \bibinfo
  {author} {\bibfnamefont {P.}~\bibnamefont {Ghosez}}, \ and\ \bibinfo {author}
  {\bibfnamefont {J.-M.}\ \bibnamefont {Triscone}},\ }\href {\doibase
  10.1146/annurev-conmatphys-062910-140445} {\bibfield  {journal} {\bibinfo
  {journal} {Annu. Rev. Conden. Ma. P.}\ }\textbf {\bibinfo {volume} {2}},\
  \bibinfo {pages} {141} (\bibinfo {year} {2011})},\ \Eprint
  {http://arxiv.org/abs/https://doi.org/10.1146/annurev-conmatphys-062910-140445}
  {https://doi.org/10.1146/annurev-conmatphys-062910-140445} \BibitemShut
  {NoStop}%
\bibitem [{\citenamefont {Glazer}(1972)}]{Glazer1972}%
  \BibitemOpen
  \bibfield  {author} {\bibinfo {author} {\bibfnamefont {A.~M.}\ \bibnamefont
  {Glazer}},\ }\href {\doibase 10.1107/S0567740872007976} {\bibfield  {journal}
  {\bibinfo  {journal} {Acta Crystallogr. B.}\ }\textbf {\bibinfo {volume}
  {28}},\ \bibinfo {pages} {3384} (\bibinfo {year} {1972})}\BibitemShut
  {NoStop}%
\bibitem [{\citenamefont {Medarde}(1997)}]{Medarde1997}%
  \BibitemOpen
  \bibfield  {author} {\bibinfo {author} {\bibfnamefont {M.~L.}\ \bibnamefont
  {Medarde}},\ }\href {http://stacks.iop.org/0953-8984/9/i=8/a=003} {\bibfield
  {journal} {\bibinfo  {journal} {J. Phys.: Condens. Matter}\ }\textbf
  {\bibinfo {volume} {9}},\ \bibinfo {pages} {1679} (\bibinfo {year}
  {1997})}\BibitemShut {NoStop}%
\bibitem [{\citenamefont {Mercy}\ \emph {et~al.}(2017)\citenamefont {Mercy},
  \citenamefont {Bieder}, \citenamefont {I\~{n}iguez},\ and\ \citenamefont
  {Ghosez}}]{Mercy2017}%
  \BibitemOpen
  \bibfield  {author} {\bibinfo {author} {\bibfnamefont {A.}~\bibnamefont
  {Mercy}}, \bibinfo {author} {\bibfnamefont {J.}~\bibnamefont {Bieder}},
  \bibinfo {author} {\bibfnamefont {J.}~\bibnamefont {I\~{n}iguez}}, \ and\
  \bibinfo {author} {\bibfnamefont {P.}~\bibnamefont {Ghosez}},\ }\href
  {https://doi.org/10.1038/s41467-017-01811-x} {\bibfield  {journal} {\bibinfo
  {journal} {Nat. Commun.}\ }\textbf {\bibinfo {volume} {8}},\ \bibinfo {pages}
  {1677} (\bibinfo {year} {2017})}\BibitemShut {NoStop}%
\bibitem [{\citenamefont {Mazin}\ \emph {et~al.}(2007)\citenamefont {Mazin},
  \citenamefont {Khomskii}, \citenamefont {Lengsdorf}, \citenamefont {Alonso},
  \citenamefont {Marshall}, \citenamefont {Ibberson}, \citenamefont
  {Podlesnyak}, \citenamefont {Mart\'{\i}nez-Lope},\ and\ \citenamefont
  {Abd-Elmeguid}}]{Mazin2007}%
  \BibitemOpen
  \bibfield  {author} {\bibinfo {author} {\bibfnamefont {I.~I.}\ \bibnamefont
  {Mazin \textit{et al.}}}  \href
  {\doibase 10.1103/PhysRevLett.98.176406} {\bibfield  {journal} {\bibinfo
  {journal} {Phys. Rev. Lett.}\ }\textbf {\bibinfo {volume} {98}},\ \bibinfo
  {pages} {176406} (\bibinfo {year} {2007})}\BibitemShut {NoStop}%
\bibitem [{\citenamefont {Park}\ \emph {et~al.}(2012)\citenamefont {Park},
  \citenamefont {Millis},\ and\ \citenamefont {Marianetti}}]{Park2012}%
  \BibitemOpen
  \bibfield  {author} {\bibinfo {author} {\bibfnamefont {H.}~\bibnamefont
  {Park}}, \bibinfo {author} {\bibfnamefont {A.~J.}\ \bibnamefont {Millis}}, \
  and\ \bibinfo {author} {\bibfnamefont {C.~A.}\ \bibnamefont {Marianetti}},\
  }\href {\doibase 10.1103/PhysRevLett.109.156402} {\bibfield  {journal}
  {\bibinfo  {journal} {Phys. Rev. Lett.}\ }\textbf {\bibinfo {volume} {109}},\
  \bibinfo {pages} {156402} (\bibinfo {year} {2012})}\BibitemShut {NoStop}%
\bibitem [{\citenamefont {Johnston}\ \emph {et~al.}(2014)\citenamefont
  {Johnston}, \citenamefont {Mukherjee}, \citenamefont {Elfimov}, \citenamefont
  {Berciu},\ and\ \citenamefont {Sawatzky}}]{Johnston2014}%
  \BibitemOpen
  \bibfield  {author} {\bibinfo {author} {\bibfnamefont {S.}~\bibnamefont
  {Johnston}}, \bibinfo {author} {\bibfnamefont {A.}~\bibnamefont {Mukherjee}},
  \bibinfo {author} {\bibfnamefont {I.}~\bibnamefont {Elfimov}}, \bibinfo
  {author} {\bibfnamefont {M.}~\bibnamefont {Berciu}}, \ and\ \bibinfo {author}
  {\bibfnamefont {G.~A.}\ \bibnamefont {Sawatzky}},\ }\href {\doibase
  10.1103/PhysRevLett.112.106404} {\bibfield  {journal} {\bibinfo  {journal}
  {Phys. Rev. Lett.}\ }\textbf {\bibinfo {volume} {112}},\ \bibinfo {pages}
  {106404} (\bibinfo {year} {2014})}\BibitemShut {NoStop}%
\bibitem [{\citenamefont {Varignon}\ \emph {et~al.}(2017)\citenamefont
  {Varignon}, \citenamefont {Grisolia}, \citenamefont {I\~{n}iguez},
  \citenamefont {Barth\'{e}l\'{e}my},\ and\ \citenamefont
  {Bibes}}]{Varignon2017a}%
  \BibitemOpen
  \bibfield  {author} {\bibinfo {author} {\bibfnamefont {J.}~\bibnamefont
  {Varignon}}, \bibinfo {author} {\bibfnamefont {M.~N.}\ \bibnamefont
  {Grisolia}}, \bibinfo {author} {\bibfnamefont {J.}~\bibnamefont
  {I\~{n}iguez}}, \bibinfo {author} {\bibfnamefont {A.}~\bibnamefont
  {Barth\'{e}l\'{e}my}}, \ and\ \bibinfo {author} {\bibfnamefont
  {M.}~\bibnamefont {Bibes}},\ }\href
  {https://doi.org/10.1038/s41535-017-0024-9} {\bibfield  {journal} {\bibinfo
  {journal} {npj Quantum Materials}\ }\textbf {\bibinfo {volume} {2}},\
  \bibinfo {pages} {21} (\bibinfo {year} {2017})}\BibitemShut {NoStop}%
\bibitem [{\citenamefont {Kimura}\ \emph {et~al.}(2003)\citenamefont {Kimura},
  \citenamefont {Ishihara}, \citenamefont {Shintani}, \citenamefont {Arima},
  \citenamefont {Takahashi}, \citenamefont {Ishizaka},\ and\ \citenamefont
  {Tokura}}]{Kimura2003}%
  \BibitemOpen
  \bibfield  {author} {\bibinfo {author} {\bibfnamefont {T.}~\bibnamefont
  {Kimura}}, \bibinfo {author} {\bibfnamefont {S.}~\bibnamefont {Ishihara}},
  \bibinfo {author} {\bibfnamefont {H.}~\bibnamefont {Shintani}}, \bibinfo
  {author} {\bibfnamefont {T.}~\bibnamefont {Arima}}, \bibinfo {author}
  {\bibfnamefont {K.~T.}\ \bibnamefont {Takahashi}}, \bibinfo {author}
  {\bibfnamefont {K.}~\bibnamefont {Ishizaka}}, \ and\ \bibinfo {author}
  {\bibfnamefont {Y.}~\bibnamefont {Tokura}},\ }\href {\doibase
  10.1103/PhysRevB.68.060403} {\bibfield  {journal} {\bibinfo  {journal} {Phys.
  Rev. B}\ }\textbf {\bibinfo {volume} {68}},\ \bibinfo {pages} {060403}
  (\bibinfo {year} {2003})}\BibitemShut {NoStop}%
\bibitem [{\citenamefont {S\'anchez}\ \emph {et~al.}(2003)\citenamefont
  {S\'anchez}, \citenamefont {Sub\'{\i}as}, \citenamefont {Garc\'{\i}a},\ and\
  \citenamefont {Blasco}}]{Sanchez2003}%
  \BibitemOpen
  \bibfield  {author} {\bibinfo {author} {\bibfnamefont {M.~C.}\ \bibnamefont
  {S\'anchez}}, \bibinfo {author} {\bibfnamefont {G.}~\bibnamefont
  {Sub\'{\i}as}}, \bibinfo {author} {\bibfnamefont {J.}~\bibnamefont
  {Garc\'{\i}a}}, \ and\ \bibinfo {author} {\bibfnamefont {J.}~\bibnamefont
  {Blasco}},\ }\href {\doibase 10.1103/PhysRevLett.90.045503} {\bibfield
  {journal} {\bibinfo  {journal} {Phys. Rev. Lett.}\ }\textbf {\bibinfo
  {volume} {90}},\ \bibinfo {pages} {045503} (\bibinfo {year}
  {2003})}\BibitemShut {NoStop}%
\bibitem [{\citenamefont {MacChesney}\ \emph {et~al.}(1965)\citenamefont
  {MacChesney}, \citenamefont {Sherwood},\ and\ \citenamefont
  {Potter}}]{MacChesney1965}%
  \BibitemOpen
  \bibfield  {author} {\bibinfo {author} {\bibfnamefont {J.~B.}\ \bibnamefont
  {MacChesney}}, \bibinfo {author} {\bibfnamefont {R.~C.}\ \bibnamefont
  {Sherwood}}, \ and\ \bibinfo {author} {\bibfnamefont {J.~F.}\ \bibnamefont
  {Potter}},\ }\href {\doibase 10.1063/1.1697052} {\bibfield  {journal}
  {\bibinfo  {journal} {J. Chem. Phys.}\ }\textbf {\bibinfo {volume} {43}},\
  \bibinfo {pages} {1907} (\bibinfo {year} {1965})}\BibitemShut {NoStop}%
\bibitem [{\citenamefont {Hayashi}\ \emph {et~al.}(2011)\citenamefont
  {Hayashi}, \citenamefont {Yamamoto}, \citenamefont {Kageyama}, \citenamefont
  {Nishi}, \citenamefont {Watanabe}, \citenamefont {Kawakami}, \citenamefont
  {Matsushita}, \citenamefont {Fujimori},\ and\ \citenamefont
  {Takano}}]{Hayashi2011}%
  \BibitemOpen
  \bibfield  {author} {\bibinfo {author} {\bibfnamefont {N.}~\bibnamefont
  {Hayashi \textit{et al.}}}} \href {\doibase
  10.1002/anie.201105276} {\bibfield  {journal} {\bibinfo  {journal} {Angew.
  Chem. Int. Edit.}\ }\textbf {\bibinfo {volume} {50}},\ \bibinfo {pages}
  {12547} (\bibinfo {year} {2011})}\BibitemShut {NoStop}%
\bibitem [{\citenamefont {Kawasaki}\ \emph {et~al.}(1998)\citenamefont
  {Kawasaki}, \citenamefont {Takano}, \citenamefont {Kanno}, \citenamefont
  {Takeda},\ and\ \citenamefont {Fujimori}}]{kawasaki1998phase}%
  \BibitemOpen
  \bibfield  {author} {\bibinfo {author} {\bibfnamefont {S.}~\bibnamefont
  {Kawasaki}}, \bibinfo {author} {\bibfnamefont {M.}~\bibnamefont {Takano}},
  \bibinfo {author} {\bibfnamefont {R.}~\bibnamefont {Kanno}}, \bibinfo
  {author} {\bibfnamefont {T.}~\bibnamefont {Takeda}}, \ and\ \bibinfo {author}
  {\bibfnamefont {A.}~\bibnamefont {Fujimori}},\ }\href@noop {} {\bibfield
  {journal} {\bibinfo  {journal} {J. Phys. Soc. Jpn.}\ }\textbf {\bibinfo
  {volume} {67}},\ \bibinfo {pages} {1529} (\bibinfo {year}
  {1998})}\BibitemShut {NoStop}%
\bibitem [{\citenamefont {Woodward}\ \emph {et~al.}(2000)\citenamefont
  {Woodward}, \citenamefont {Cox}, \citenamefont {Moshopoulou}, \citenamefont
  {Sleight},\ and\ \citenamefont {Morimoto}}]{Woodward2000}%
  \BibitemOpen
  \bibfield  {author} {\bibinfo {author} {\bibfnamefont {P.~M.}\ \bibnamefont
  {Woodward}}, \bibinfo {author} {\bibfnamefont {D.~E.}\ \bibnamefont {Cox}},
  \bibinfo {author} {\bibfnamefont {E.}~\bibnamefont {Moshopoulou}}, \bibinfo
  {author} {\bibfnamefont {A.~W.}\ \bibnamefont {Sleight}}, \ and\ \bibinfo
  {author} {\bibfnamefont {S.}~\bibnamefont {Morimoto}},\ }\href {\doibase
  10.1103/PhysRevB.62.844} {\bibfield  {journal} {\bibinfo  {journal} {Phys.
  Rev. B}\ }\textbf {\bibinfo {volume} {62}},\ \bibinfo {pages} {844} (\bibinfo
  {year} {2000})}\BibitemShut {NoStop}%
\bibitem [{\citenamefont {Akao}\ \emph {et~al.}(2003)\citenamefont {Akao},
  \citenamefont {Azuma}, \citenamefont {Usuda}, \citenamefont {Nishihata},
  \citenamefont {Mizuki}, \citenamefont {Hamada}, \citenamefont {Hayashi},
  \citenamefont {Terashima},\ and\ \citenamefont {Takano}}]{Akao2003}%
  \BibitemOpen
  \bibfield  {author} {\bibinfo {author} {\bibfnamefont {T.}~\bibnamefont
  {Akao\textit{et al.}}}}} \href {\doibase 10.1103/PhysRevLett.91.156405}
  {\bibfield  {journal} {\bibinfo  {journal} {Phys. Rev. Lett.}\ }\textbf
  {\bibinfo {volume} {91}},\ \bibinfo {pages} {156405} (\bibinfo {year}
  {2003})}\BibitemShut {NoStop}%
\bibitem [{\citenamefont {Matsuno}\ \emph {et~al.}(2002)\citenamefont
  {Matsuno}, \citenamefont {Mizokawa}, \citenamefont {Fujimori}, \citenamefont
  {Takeda}, \citenamefont {Kawasaki},\ and\ \citenamefont
  {Takano}}]{Matsuno2002}%
  \BibitemOpen
  \bibfield  {author} {\bibinfo {author} {\bibfnamefont {J.}~\bibnamefont
  {Matsuno}}, \bibinfo {author} {\bibfnamefont {T.}~\bibnamefont {Mizokawa}},
  \bibinfo {author} {\bibfnamefont {A.}~\bibnamefont {Fujimori}}, \bibinfo
  {author} {\bibfnamefont {Y.}~\bibnamefont {Takeda}}, \bibinfo {author}
  {\bibfnamefont {S.}~\bibnamefont {Kawasaki}}, \ and\ \bibinfo {author}
  {\bibfnamefont {M.}~\bibnamefont {Takano}},\ }\href {\doibase
  10.1103/PhysRevB.66.193103} {\bibfield  {journal} {\bibinfo  {journal} {Phys.
  Rev. B}\ }\textbf {\bibinfo {volume} {66}},\ \bibinfo {pages} {193103}
  (\bibinfo {year} {2002})}\BibitemShut {NoStop}%
\bibitem [{\citenamefont {Ghosh}\ \emph {et~al.}(2005)\citenamefont {Ghosh},
  \citenamefont {Kamaraju}, \citenamefont {Seto}, \citenamefont {Fujimori},
  \citenamefont {Takeda}, \citenamefont {Ishiwata}, \citenamefont {Kawasaki},
  \citenamefont {Azuma}, \citenamefont {Takano},\ and\ \citenamefont
  {Sood}}]{Ghosh2005}%
  \BibitemOpen
  \bibfield  {author} {\bibinfo {author} {\bibfnamefont {S.}~\bibnamefont
  {Ghosh \textit{et al.}}}}, \href {\doibase 10.1103/PhysRevB.71.245110}
  {\bibfield  {journal} {\bibinfo  {journal} {Phys. Rev. B}\ }\textbf {\bibinfo
  {volume} {71}},\ \bibinfo {pages} {245110} (\bibinfo {year}
  {2005})}\BibitemShut {NoStop}%
\bibitem [{\citenamefont {Cammarata}\ and\ \citenamefont
  {Rondinelli}(2012)}]{Cammarata2012}%
  \BibitemOpen
  \bibfield  {author} {\bibinfo {author} {\bibfnamefont {A.}~\bibnamefont
  {Cammarata}}\ and\ \bibinfo {author} {\bibfnamefont {J.~M.}\ \bibnamefont
  {Rondinelli}},\ }\href {\doibase 10.1103/PhysRevB.86.195144} {\bibfield
  {journal} {\bibinfo  {journal} {Phys. Rev. B}\ }\textbf {\bibinfo {volume}
  {86}},\ \bibinfo {pages} {195144} (\bibinfo {year} {2012})}\BibitemShut
  {NoStop}%
\bibitem [{\citenamefont {Rogge}\ \emph {et~al.}(2018)\citenamefont {Rogge},
  \citenamefont {Chandrasena}, \citenamefont {Cammarata}, \citenamefont
  {Green}, \citenamefont {Shafer}, \citenamefont {Lefler}, \citenamefont
  {Huon}, \citenamefont {Arab}, \citenamefont {Arenholz}, \citenamefont {Lee},
  \citenamefont {Lee}, \citenamefont {Nem\ifmmode~\check{s}\else
  \v{s}\fi{}\'ak}, \citenamefont {Rondinelli}, \citenamefont {Gray},\ and\
  \citenamefont {May}}]{Rogge2018}%
  \BibitemOpen
  \bibfield  {author} {\bibinfo {author} {\bibfnamefont {P.~C.}\ \bibnamefont
  {Rogge \textit{et al.}}},\ }\href {\doibase
  10.1103/PhysRevMaterials.2.015002} {\bibfield  {journal} {\bibinfo  {journal}
  {Phys. Rev. Materials}\ }\textbf {\bibinfo {volume} {2}},\ \bibinfo {pages}
  {015002} (\bibinfo {year} {2018})}\BibitemShut {NoStop}%
\bibitem [{\citenamefont {Kresse}\ and\ \citenamefont
  {Hafner}(1993)}]{Kresse1993}%
  \BibitemOpen
  \bibfield  {author} {\bibinfo {author} {\bibfnamefont {G.}~\bibnamefont
  {Kresse}}\ and\ \bibinfo {author} {\bibfnamefont {J.}~\bibnamefont
  {Hafner}},\ }\href {\doibase 10.1103/PhysRevB.47.558} {\bibfield  {journal}
  {\bibinfo  {journal} {Phys. Rev. B}\ }\textbf {\bibinfo {volume} {47}},\
  \bibinfo {pages} {558} (\bibinfo {year} {1993})}\BibitemShut {NoStop}%
\bibitem [{\citenamefont {Bl\"ochl}(1994)}]{Bloechl1994}%
  \BibitemOpen
  \bibfield  {author} {\bibinfo {author} {\bibfnamefont {P.~E.}\ \bibnamefont
  {Bl\"ochl}},\ }\href {\doibase 10.1103/PhysRevB.50.17953} {\bibfield
  {journal} {\bibinfo  {journal} {Phys. Rev. B}\ }\textbf {\bibinfo {volume}
  {50}},\ \bibinfo {pages} {17953} (\bibinfo {year} {1994})}\BibitemShut
  {NoStop}%
\bibitem [{\citenamefont {Perdew}\ \emph {et~al.}(2008)\citenamefont {Perdew},
  \citenamefont {Ruzsinszky}, \citenamefont {Csonka}, \citenamefont {Vydrov},
  \citenamefont {Scuseria}, \citenamefont {Constantin}, \citenamefont {Zhou},\
  and\ \citenamefont {Burke}}]{Perdew2008}%
  \BibitemOpen
  \bibfield  {author} {\bibinfo {author} {\bibfnamefont {J.~P.}\ \bibnamefont
  {Perdew}}, \bibinfo {author} {\bibfnamefont {A.}~\bibnamefont {Ruzsinszky}},
  \bibinfo {author} {\bibfnamefont {G.~I.}\ \bibnamefont {Csonka}}, \bibinfo
  {author} {\bibfnamefont {O.~A.}\ \bibnamefont {Vydrov}}, \bibinfo {author}
  {\bibfnamefont {G.~E.}\ \bibnamefont {Scuseria}}, \bibinfo {author}
  {\bibfnamefont {L.~A.}\ \bibnamefont {Constantin}}, \bibinfo {author}
  {\bibfnamefont {X.}~\bibnamefont {Zhou}}, \ and\ \bibinfo {author}
  {\bibfnamefont {K.}~\bibnamefont {Burke}},\ }\href {\doibase
  10.1103/PhysRevLett.100.136406} {\bibfield  {journal} {\bibinfo  {journal}
  {Phys. Rev. Lett.}\ }\textbf {\bibinfo {volume} {100}},\ \bibinfo {pages}
  {136406} (\bibinfo {year} {2008})}\BibitemShut {NoStop}%
\bibitem [{\citenamefont {Anisimov}\ \emph {et~al.}(1997)\citenamefont
  {Anisimov}, \citenamefont {Aryasetiawan},\ and\ \citenamefont
  {Lichtenstein}}]{Anisimov1997}%
  \BibitemOpen
  \bibfield  {author} {\bibinfo {author} {\bibfnamefont {V.~I.}\ \bibnamefont
  {Anisimov}}, \bibinfo {author} {\bibfnamefont {F.}~\bibnamefont
  {Aryasetiawan}}, \ and\ \bibinfo {author} {\bibfnamefont {A.~I.}\
et  \bibnamefont {Lichtenstein}},\ }\href
  {http://stacks.iop.org/0953-8984/9/i=4/a=002} {\bibfield  {journal} {\bibinfo
   {journal} {J. Phys.: Condens. Matter}\ }\textbf {\bibinfo {volume} {9}},\
  \bibinfo {pages} {767} (\bibinfo {year} {1997})}\BibitemShut {NoStop}%
\bibitem [{\citenamefont {Monkhorst}\ and\ \citenamefont
  {Pack}(1976)}]{PhysRevB.13.5188}%
  \BibitemOpen
  \bibfield  {author} {\bibinfo {author} {\bibfnamefont {H.~J.}\ \bibnamefont
  {Monkhorst}}\ and\ \bibinfo {author} {\bibfnamefont {J.~D.}\ \bibnamefont
  {Pack}},\ }\href {\doibase 10.1103/PhysRevB.13.5188} {\bibfield  {journal}
  {\bibinfo  {journal} {Phys. Rev. B}\ }\textbf {\bibinfo {volume} {13}},\
  \bibinfo {pages} {5188} (\bibinfo {year} {1976})}\BibitemShut {NoStop}%
\bibitem [{\citenamefont {Campbell}\ \emph {et~al.}(2006)\citenamefont
  {Campbell}, \citenamefont {Stokes}, \citenamefont {Tanner},\ and\
  \citenamefont {Hatch}}]{Campbell2006}%
  \BibitemOpen
  \bibfield  {author} {\bibinfo {author} {\bibfnamefont {B.~J.}\ \bibnamefont
  {Campbell}}, \bibinfo {author} {\bibfnamefont {H.~T.}\ \bibnamefont
  {Stokes}}, \bibinfo {author} {\bibfnamefont {D.~E.}\ \bibnamefont {Tanner}},
  \ and\ \bibinfo {author} {\bibfnamefont {D.~M.}\ \bibnamefont {Hatch}},\
  }\href {\doibase 10.1107/S0021889806014075} {\bibfield  {journal} {\bibinfo
  {journal} {J. Appl. Crystallogr.}\ }\textbf {\bibinfo {volume} {39}},\
  \bibinfo {pages} {607} (\bibinfo {year} {2006})}\BibitemShut {NoStop}%
\bibitem [{\citenamefont {Holakovsk\'{y}}(1973)}]{Holakovsky1973}%
  \BibitemOpen
  \bibfield  {author} {\bibinfo {author} {\bibfnamefont {J.}~\bibnamefont
  {Holakovsk\'{y}}},\ }\href {\doibase 10.1002/pssb.2220560224} {\bibfield
  {journal} {\bibinfo  {journal} {physica status solidi (b)}\ }\textbf
  {\bibinfo {volume} {56}},\ \bibinfo {pages} {615} (\bibinfo {year}
  {1973})}\BibitemShut {NoStop}%
\bibitem [{\citenamefont {Takeda}\ \emph {et~al.}(2000)\citenamefont {Takeda},
  \citenamefont {Kanno}, \citenamefont {Kawamoto}, \citenamefont {Takano},
  \citenamefont {Kawasaki}, \citenamefont {Kamiyama},\ and\ \citenamefont
  {Izumi}}]{Takeda2000}%
  \BibitemOpen
  \bibfield  {author} {\bibinfo {author} {\bibfnamefont {T.}~\bibnamefont
  {Takeda}}, \bibinfo {author} {\bibfnamefont {R.}~\bibnamefont {Kanno}},
  \bibinfo {author} {\bibfnamefont {Y.}~\bibnamefont {Kawamoto}}, \bibinfo
  {author} {\bibfnamefont {M.}~\bibnamefont {Takano}}, \bibinfo {author}
  {\bibfnamefont {S.}~\bibnamefont {Kawasaki}}, \bibinfo {author}
  {\bibfnamefont {T.}~\bibnamefont {Kamiyama}}, \ and\ \bibinfo {author}
  {\bibfnamefont {F.}~\bibnamefont {Izumi}},\ }\href {\doibase
  10.1016/S1293-2558(00)01088-8} {\bibfield  {journal} {\bibinfo  {journal}
  {Solid State Sci.}\ }\textbf {\bibinfo {volume} {2}},\ \bibinfo {pages} {673}
  (\bibinfo {year} {2000})}\BibitemShut {NoStop}%
\bibitem [{\citenamefont {Whangbo}\ \emph {et~al.}(2002)\citenamefont
  {Whangbo}, \citenamefont {Koo}, \citenamefont {Villesuzanne},\ and\
  \citenamefont {Pouchard}}]{Whangbo2002}%
  \BibitemOpen
  \bibfield  {author} {\bibinfo {author} {\bibfnamefont {M.-H.}\ \bibnamefont
  {Whangbo}}, \bibinfo {author} {\bibfnamefont {H.-J.}\ \bibnamefont {Koo}},
  \bibinfo {author} {\bibfnamefont {A.}~\bibnamefont {Villesuzanne}}, \ and\
  \bibinfo {author} {\bibfnamefont {M.}~\bibnamefont {Pouchard}},\ }\href
  {\doibase 10.1021/ic0110427} {\bibfield  {journal} {\bibinfo  {journal}
  {Inorg. Chem.}\ }\textbf {\bibinfo {volume} {41}},\ \bibinfo {pages} {1920}
  (\bibinfo {year} {2002})},\ \bibinfo {note} {pMID: 11925189},\ \Eprint
  {http://arxiv.org/abs/https://doi.org/10.1021/ic0110427}
  {https://doi.org/10.1021/ic0110427} \BibitemShut {NoStop}%
\bibitem [{\citenamefont {Varignon}\ \emph {et~al.}(2015)\citenamefont
  {Varignon}, \citenamefont {Bristowe}, \citenamefont {Bousquet},\ and\
  \citenamefont {Ghosez}}]{Varignon2015}%
  \BibitemOpen
  \bibfield  {author} {\bibinfo {author} {\bibfnamefont {J.}~\bibnamefont
  {Varignon}}, \bibinfo {author} {\bibfnamefont {N.~C.}\ \bibnamefont
  {Bristowe}}, \bibinfo {author} {\bibfnamefont {E.}~\bibnamefont {Bousquet}},
  \ and\ \bibinfo {author} {\bibfnamefont {P.}~\bibnamefont {Ghosez}},\ }\href
  {http://www.ncbi.nlm.nih.gov/pmc/articles/PMC4612717/} {\bibfield  {journal}
  {\bibinfo  {journal} {Sci. Rep.}\ }\textbf {\bibinfo {volume} {5}},\ \bibinfo
  {pages} {15364} (\bibinfo {year} {2015})}\BibitemShut {NoStop}%
\bibitem [{\citenamefont {Carpenter}\ and\ \citenamefont
  {Howard}(2009)}]{Carpenter2009}%
  \BibitemOpen
  \bibfield  {author} {\bibinfo {author} {\bibfnamefont {M.~A.}\ \bibnamefont
  {Carpenter}}\ and\ \bibinfo {author} {\bibfnamefont {C.~J.}\ \bibnamefont
  {Howard}},\ }\href {\doibase 10.1107/S0108768109000974} {\bibfield  {journal}
  {\bibinfo  {journal} {Acta Crystallogr. B.}\ }\textbf {\bibinfo {volume}
  {65}},\ \bibinfo {pages} {134} (\bibinfo {year} {2009})}\BibitemShut
  {NoStop}%
\bibitem [{\citenamefont {Yin}\ \emph {et~al.}(2006)\citenamefont {Yin},
  \citenamefont {Volja},\ and\ \citenamefont {Ku}}]{Yin2006}%
  \BibitemOpen
  \bibfield  {author} {\bibinfo {author} {\bibfnamefont {W.-G.}\ \bibnamefont
  {Yin}}, \bibinfo {author} {\bibfnamefont {D.}~\bibnamefont {Volja}}, \ and\
  \bibinfo {author} {\bibfnamefont {W.}~\bibnamefont {Ku}},\ }\href {\doibase
  10.1103/PhysRevLett.96.116405} {\bibfield  {journal} {\bibinfo  {journal}
  {Phys. Rev. Lett.}\ }\textbf {\bibinfo {volume} {96}},\ \bibinfo {pages}
  {116405} (\bibinfo {year} {2006})}\BibitemShut {NoStop}%
\bibitem [{\citenamefont {He}\ and\ \citenamefont {Millis}(2015)}]{He2015}%
  \BibitemOpen
  \bibfield  {author} {\bibinfo {author} {\bibfnamefont {Z.}~\bibnamefont
  {He}}\ and\ \bibinfo {author} {\bibfnamefont {A.~J.}\ \bibnamefont
  {Millis}},\ }\href {\doibase 10.1103/PhysRevB.91.195138} {\bibfield
  {journal} {\bibinfo  {journal} {Phys. Rev. B}\ }\textbf {\bibinfo {volume}
  {91}},\ \bibinfo {pages} {195138} (\bibinfo {year} {2015})}\BibitemShut
  {NoStop}%
\bibitem [{\citenamefont {35}}]{foot}%
  \BibitemOpen
  \bibfield  {author} {\bibinfo {author} {\bibfnamefont {See supplementary material at \textit{URL} for details about the determination of (U$|$J) calculation parameters, supporting calculations on Sr/BeFeO$_3$ solutions, and a deeper analysis on the assymetry of the $Q_{JT}$ energy surface}}}\BibitemShut {NoStop}%
\end{thebibliography}
\end{document}